\documentclass[a4paper,11pt,fleqn]{article}
\pdfoutput=1

\usepackage{amsmath}
\usepackage{amssymb}
\usepackage{graphicx}
\usepackage{bbm}
\usepackage{cite}

\usepackage[DIV12]{typearea}
\usepackage{amsfonts}
\usepackage{amscd}
\usepackage[footnotesize]{caption}
\usepackage{mathrsfs}
\usepackage{color}

\newcommand{\be}{\begin{equation}}  
\newcommand{\ee}{\end{equation}}  
\newcommand{\ol}[1]{\overline{#1}}
\newcommand{\hc}{+\,\mathrm{h.c.}}
\newcommand{\vev}[1]{\langle #1 \rangle}

\newcommand{\U}[1]{\ensuremath{\mathrm{U}(#1)}}
\newcommand{\into}{\ensuremath{\;\rightarrow\;}}
\newcommand{\tr}{\operatorname{tr}}

\newcommand{\wh}[1]{\ensuremath{\widehat{#1}}}

\title{
\vspace{-4.5ex}
{\normalsize \raggedright
DESY 13--174\\
November 2013\\[10ex]
}
\textbf{A low Fermi scale \\ from a simple gaugino-scalar mass relation} 
\vspace{2ex}
}

\author{\large F.~Br\"ummer$^{a,b}$ and W.~Buchm\"uller$^{b}$\\[1ex]
\textit{\normalsize $^a$ Scuola Internazionale Superiore di Studi Avanzati SISSA/ISAS,}\\
\textit{\normalsize Via Bonomea 265, I-34136 Trieste, Italy}\\
\textit{\normalsize $^b$ Deutsches Elektronen-Synchrotron DESY,}\\
\textit{\normalsize Notkestra\ss e 85, D-22607 Hamburg, Germany}\\[1ex]
\vspace{3ex}
}

\date{}

\begin{document}

\maketitle

\begin{abstract}
\noindent 
In supersymmetric extensions of the Standard Model, the Fermi scale of
electroweak symmetry breaking is determined by the pattern of
supersymmetry breaking. We present an example, motivated by a higher-dimensional
GUT model, where a particular mass relation between the gauginos, third-generation 
squarks and Higgs fields of the MSSM leads to a Fermi scale smaller than 
the soft mass scale. This is in agreement with the measured Higgs
boson mass. The $\mu$ parameter is generated independently of supersymmetry
breaking, however the $\mu$ problem becomes less acute due to the little 
hierarchy between the soft mass scale and the Fermi scale as we will argue. 
The resulting superparticle mass spectra depend on the localization of
quark and lepton fields in higher dimensions. In one case, the
squarks of the first two  generations as well as the gauginos and higgsinos
can be in the range of the LHC. Alternatively, only the higgsinos may
be accessible at colliders. The lightest superparticle is the gravitino. 

\end{abstract}

\section{Introduction}

The unification of gauge couplings and the prediction of viable dark matter
candidates provides a strong theoretical motivation for supersymmetric
extensions of the Standard Model with TeV superparticle masses 
\cite{Dimopoulos:1981zb,Ibanez:1981yh,Dimopoulos:1981yj}.
So far searches for heavy superparticles at the Large Hadron Collider (LHC)
have only led to lower bounds on scalar quark and gluino masses
of about 1--2~TeV \cite{Aad:2013wta,Chatrchyan:2013lya}. 
On the other hand, the discovery of a 126~GeV Higgs 
boson \cite{Aad:2012tfa,Chatrchyan:2012ufa} allows, without or with supersymmetry, for an 
extrapolation of the Standard Model up to the scale of grand unification.

The Higgs boson mass is consistent with the mass range predicted by the
minimal supersymmetric standard model (MSSM). However, since the
Higgs mass significantly exceeds its tree-level upper bound of 91~GeV,
quantum corrections are large, which generically requires multi-TeV scalar
masses. This raises the question why the Fermi scale, the expectation value
of the Higgs field, 
$\langle H\rangle = (\sqrt{2}G_F)^{-1/2} = 246~\mathrm{GeV}$, is much 
smaller than the scale of supersymmetry breaking, and the required fine-tuning
of seemingly unrelated parameters is often considered as unnatural. 
Possible answers to this question invoke the anthropic principle and the
string landscape, as in split supersymmetry \cite{ArkaniHamed:2004fb},
the focus point idea \cite{Chan:1997bi}, or similar accidental 
cancellations between non-universal gaugino and scalar masses 
at the grand unification scale \cite{Abe:2007kf}. The naturalness
problem might also be solved in a non-minimal extension of the MSSM with
additional sub-TeV degrees of freedom (for instance, the NMSSM, reviewed
in \cite{Ellwanger:2009dp}), or through non-decoupling effects such as in
\cite{Batra:2003nj}.\footnote{See \cite{Craig:2013cxa} for a 
recent review of naturalness in supersymmetry in the light of the first LHC run.}

In this note we restrict ourselves to the MSSM, and attempt to answer a
question which is intimately connected with the naturalness problem: Is there 
a well motivated and simple set of boundary conditions for the GUT-scale soft 
terms which favours a `little hierarchy' between the soft and the electroweak 
scale? And, what can we expect from this soft mass pattern for the upcoming 
second LHC run?

Our main findings can be summarized as follows. Since the $\mu$ parameter
of the MSSM can be generated independently of supersymmetry breaking, it
is technically natural to choose it smaller than the typical soft SUSY
breaking parameters, say, of the order of the electroweak scale. Usually, 
explaining why $\mu$ should be of the order of the soft 
masses is a well known challenge (the `$\mu$ problem') in SUSY model
building. Here, as we will argue, the $\mu$ problem becomes less severe once
one accepts a little hierarchy. To obtain proper electroweak symmetry 
breaking at large $\tan\beta$, the loop-corrected up-type Higgs soft mass needs 
to be of the same order as $\mu$ at the scale where the MSSM is matched to the 
Standard Model, requiring an accidental cancellation between the 
tree-level and radiative contributions to this parameter. We identify a simple 
soft mass pattern which suggests this cancellation, and which is motivated by
a six-dimensional GUT model (although we expect that there are other models
that can lead to the same pattern). Within this model, we obtain an estimate for 
the possible range of the gluino mass, which will be partly probed at LHC-14. 
Squarks and sleptons may also be within reach, and by construction there are 
higgsino-like charginos and neutralinos with electroweak-scale masses which
can be discovered at a linear collider.

Note that we are not claiming to solve the fine-tuning problem: The fine-tuning 
in our model is as large as one would expect from a generic MSSM-type model 
without large contributions to the lightest Higgs mass from stop mixing, 
i.e.~at the permille level. Our model predicts the relevant soft terms only
up to factors of order one, and while the predicted pattern non-trivially 
allows for a little hierarchy, these unknown factors still need to be tuned
in order to actually realize it. We anticipate that fully understanding the 
origin of the cancellations involved will require a better understanding 
of the complete UV theory.

\section{Electroweak symmetry breaking with a little hierarchy}\label{littleh}

\subsubsection*{Matching the MSSM to the Standard Model}

The scalar potential for the MSSM Higgs fields depends on the higgsino
mass $\mu$, which is a parameter of the superpotential, and the soft
supersymmetry breaking parameters $m_{H_u}^2$, $m_{H_d}^2$ and $B\mu$,  
\be\begin{split}\label{MSSM}
V = &\left(m_{H_u}^2+|\mu|^2\right)H_u^\dagger H_u
+  \left(m_{H_d}^2+|\mu|^2\right)H_d^\dagger H_d
+  B\mu \left(H_u^T i\sigma_2 H_d + {\rm c.c.}\right) \\
&+ \frac{1}{8}\left(g^2 +g'^2\right)\left(H_u^\dagger H_u - H_d^\dagger H_d\right)^2
+ \frac{1}{2}\,g^2\ H_u^\dagger H_d H_d^\dagger H_u \ .
\end{split}\ee
Our starting assumption is that the only scalar with an electroweak-scale
mass is the lightest Higgs, while all others (in particular the remaining Higgs
bosons) are much heavier. In this so-called decoupling limit the Higgs vacuum
expectation value (vev) is 
approximately aligned with the lightest mass eigenstate. It is convenient 
to work with the fields $H$ and $H'$ defined by\footnote{In the usual notation for the tree-level
mass eigenstates, would-be Goldstone bosons, and mixing angles (see 
e.g.~\cite{Martin:1997ns}) this corresponds 
to $H=\left(G^+,\;v+(h^0+iG^0)/\sqrt{2}\right)^T$, 
$H'=\left((H^0+iA^0)/\sqrt{2},\;H^{+*}\right)^T$, and $\alpha=\beta-\pi/2$.
} 
\be
H_u = \sin\beta\ H + \cos\beta\ i\sigma_2 H'^* \ , \quad
H_d = \cos\beta i\sigma_2 H^* + \sin\beta H' \ , \\
\ee
with
\be\label{t2b}
\tan{2\beta} = \frac{2B\mu}{m_{H_u}^2 - m_{H_d}^2} \ ,
\ee
such that the quadratic part of the potential is
diagonal in the new fields:
\be\begin{split}
V =\ &m^2 H^\dagger H + m'^2 H'^\dagger H' \\
&+ \frac{1}{8}\left(g^2 +g'^2\right)\left(\cos{2\beta}\left(H^\dagger H -  H'^\dagger H'\right)
- \sin{2\beta} \left(H^T i\sigma_2 H' + {\rm c.c.}\right)\right)^2 \\
&+ \frac{1}{2}g^2\ H^\dagger H' H'^\dagger H \ , 
\end{split}\ee
where
\begin{align}
m^2 &= |\mu|^2 + m_{H_u}^2\sin^2\beta + m_{H_d}^2\cos^2\beta -
B\mu\sin{2\beta}\ ,\\
m'^2 &= |\mu|^2 + m_{H_u}^2\cos^2\beta + m_{H_d}^2\sin^2\beta +
B\mu\sin{2\beta}\ \ .
\end{align}

Within the MSSM the measured mass of the lightest Higgs boson requires large
radiative corrections from heavy stop squarks. 
Therefore we take the scale $M_S = (m_{\tilde t_1} m_{\tilde t_2})^{1/2}$
to be much larger than the electroweak scale, of the order of several TeV. At the
scale $M_S$ the MSSM is matched to the Standard Model with scalar potential
\be
V =\ m^2 H^\dagger H + \frac{1}{2}\lambda\left( H^\dagger H\right)^2 \ ,
\ee
where
\be\label{quartic}
\left.\lambda\right|_{M_S} =\ \frac{1}{4}\left.\left(g^2
    +g'^2\right)\cos^2{2\beta}\right|_{M_S} \ .
\ee
The Higgs mass parameter $m^2$ encodes the prediction
for the electroweak scale, $v^2=-m^2/\lambda$, with $\lambda$
being ${\cal O}(1)$.

\subsubsection*{Conditions for a little hierarchy}

When keeping the electroweak scale fixed, the tree-level contribution to the
lightest Higgs mass is maximized at large $\tan\beta$ (since 
in that limit $|\cos 2\beta|\into 1$ in Eq.~\eqref{quartic}), approaching its
limit value of $m_Z=91$ GeV. The region of at least moderately large 
$\tan\beta\gtrsim 10$ is therefore 
favoured by the large observed Higgs mass of 126 GeV, with the discrepancy
accounted for by radiative corrections. The Standard Model-like Higgs field 
$H$ is then predominantly $H_u$.

By Eq.~\eqref{t2b}, using $\tan 2\beta=2/(\cot\beta-\tan\beta)$, large
$\tan\beta$ implies 
\be\label{smallBmu}
B\mu\ll m_{H_d}^2
\ee
in the generic case that $\left|m_{H_u}^2\right|\lesssim m_{H_d}^2$.\footnote{We do not consider exceptionally small values for $m_{H_d}^2$,
which could occur in exotic mediation schemes or be induced by RG running
at large $y_b$ (i.e.~extremely large $\tan\beta \gtrsim 40$). Some more details
about the running of $m_{H_d}^2$ are given below.}
In the following we will take the $\mu$ parameter to be generated independently 
of supersymmetry breaking.  Since $\mu$ and $B\mu$ are both governed 
by a Peccei-Quinn symmetry, 
unless $B\mu$ is merely accidentally small due to radiative corrections,
the reason underlying relation \eqref{smallBmu} is that the effective symmetry breaking 
scale is below the soft mass scale, as will be discussed in more detail momentarily. 
In that case also $\mu$ is small:
\be\label{smallmu}
|\mu|^2\ll m_{H_d}^2\ .
\ee
Furthermore, since at large $\tan\beta$ we have\footnote{The last 
term in Eq.~\eqref{Fermi} is often neglected. However, in the
case of a large matching scale $M_S$ it is generally important, even
for large values of $\tan\beta$.}
\be
m^2 \simeq |\mu|^2 + m_{H_u}^2 +\frac{m_{H_u}^2-m_{H_d}^2}{\tan^2\beta} \ , \label{Fermi}
\ee
a little hierarchy requires that $m_{H_u}^2$ is small,
\be
\left|m_{H_u}^2\right|\ll m_{H_d}^2\ .\label{smallmHu}
\ee
Together with Eq.~\eqref{t2b} this implies
\be
\tan{\beta} \simeq \frac{m_{H_d}^2}{B\mu} \ . \label{Bmu}
\ee
Relations \eqref{smallBmu}, \eqref{smallmu} and \eqref{smallmHu}
are thus necessary to obtain a Fermi scale much smaller than the soft
mass scale, assuming that $\tan\beta$ is at least moderately large 
and that $\mu$ and $B\mu$ are connected. We now proceed to discuss the
possible origins of these conditions.

\subsubsection*{Why should $m_{H_u}^2$ be small?}

Choosing $\mu$ and $B\mu$ small is technically
natural, and this choice is radiatively stable.
By contrast, radiative corrections to the Higgs soft masses are sizeable. 
In particular, as is well known, no symmetry protects $m_{H_u}^2$
from loop corrections due to the large top Yukawa coupling. Condition
\eqref{smallmHu} is technically unnatural, which is a manifestation of the usual
fine-tuning problem in the MSSM. It requires large cancellations between
the radiative contributions and the tree-level value of $m_{H_u}^2$. Let us
discuss these in some more detail.

When considering models whose fundamental parameters are defined
at the GUT scale, then the Higgs
potential will receive large logarithmically enhanced quantum corrections, 
which need to be resummed using the MSSM renormalization group equations. 
In addition, there are finite corrections at the matching scale $M_S$
which we cannot neglect.

Turning first to the renormalization group running, 
the tree-level RG-improved Higgs potential at $M_S$ can be expressed as a function of the running Higgs mass parameters and of the running gauge couplings. The Higgs mass parameters at the scale $M_S$ depend on their GUT-scale values, but also on the GUT-scale soft masses of all fields with sizeable couplings to the Higgs sector. These are the third-generation scalars and the gauginos (with the gluino entering because of its large coupling to the stops and sbottoms). We find for $\tan\beta=15$
\be\begin{split}\label{mHusq2}
\left.m_{H_u}^2\right|_{M_S}=&-\left\{\begin{array}{c}1.09 \\ 1.13 \\ 1.18\end{array}\right\}\wh {M_3}^2-\left\{\begin{array}{c}0.10 \\ 0.11 \\ 0.11\end{array}\right\}\wh M_3\wh M_2+0.22\,\wh M_2^2+0.26\,\wh M_3\wh{A}_t+0.07\,\wh M_2\wh{A}_t\\
&-0.12\,\wh{A}_{t}^2+\left\{\begin{array}{c}0.67 \\ 0.67 \\ 0.66\end{array}\right\}{\wh m_{H_u}}^2-0.24\,{\wh m_{U_3}}^2-\left\{\begin{array}{c}0.33 \\ 0.33 \\ 0.34\end{array}\right\}{\wh m_{Q_3}}^2\ ,\\
&\text{for } M_S=\left\{\begin{array}{c}6.5 \\ 5 \\ 3.5\end{array}\right\}\text{ TeV}\,.
\end{split}
\ee
Here the hatted quantities on the RHS denote GUT-scale soft
parameters, with $\wh A_t$ normalized to the top Yukawa coupling. We
have taken the GUT scale to be fixed at $M_{\rm GUT}=1.5\times 10^{16}$ GeV, and
omitted all terms with coefficients smaller than $0.05$. The
coefficients are largely insensitive to $\tan\beta$, as long as
$\tan\beta\gtrsim 10$; for instance, for $\tan\beta=30$ the
coefficients of the $\wh M_3^2$ term are $-\{1.07;\,1.11;\,1.16\}$ and
all other coefficients differ from Eq.~\eqref{mHusq2} at most by
$0.01$. Another source of uncertainty is the experimental uncertainty
in the top mass. We have checked that the uncertainty obtained from
varying $m_t$ by 1$\sigma$ around its central value of $173.2$ GeV is
of similar order, changing the coefficients at most by 
$0.01$. The GUT-scale values for Yukawa and gauge couplings have been
obtained using the two-loop RG code {\tt SOFTSUSY} \cite{Allanach:2001kg}.

With the assumptions of large $\tan\beta$ and of negligible stop mixing at the GUT scale, i.e.~negligible $\wh A_t$, the matching scale is in principle rather sharply determined by the lightest Higgs mass $m_{h^0}$. This is because the radiative corrections to $m_{h^0}$ depend mainly on the stop masses (and on the RG-induced stop mixing parameter at the TeV scale). In practice however there is still a large uncertainty, partly because of the uncertainty in $y_t$, but mostly because of the theory uncertainty in computing $m_{h^0}$ from a given soft mass spectrum. We have chosen $M_S=5\pm 1.5$ TeV, which is in good accordance with the two-loop spectrum codes {\tt SOFTSUSY}, {\tt SuSpect} \cite{Djouadi:2002ze} and {\tt FeynHiggs} \cite{Heinemeyer:1998yj} and also compatible with the three-loop analysis in \cite{Feng:2013tvd} which is based on the {\tt H3M} code \cite{Kant:2010tf}.

The equivalent of Eq.~\eqref{mHusq2} for $m_{H_d}^2$ reads
\be\begin{split}\label{mHdsq2}
\left.m_{H_d}^2\right|_{M_S}=&-\left\{\begin{array}{c}0.06 \\ 0.07 \\ 0.07\end{array}\right\}\wh {M}_3^2+\left\{\begin{array}{c}0.37 \\ 0.38 \\ 0.38\end{array}\right\}\wh {M_2}^2
+\left\{\begin{array}{c}0.95 \\ 0.95 \\ 0.94\end{array}\right\}\,{\wh m_{H_d}}^2
-0.06\,{\wh m_{U_3}}^2\ ,\\
&\text{for } M_S=\left\{\begin{array}{c}6.5 \\ 5 \\ 3.5\end{array}\right\}\text{ TeV}\,.
\end{split}
\ee
While the coefficients in Eq.~\eqref{mHdsq2} show a more pronounced
$\tan\beta$-dependence, the overall running of $m_{H_d}^2$ remains moderate 
in the range $10\lesssim\tan\beta\lesssim 40$.
In this region $\left.m_{H_d}\right|_{M_S}$ is therefore of the order of $\wh m_{H_d}$, which is
generically  of the order $M_S$.

In addition to the RG running of the tree-level
  parameters there are important finite corrections to the Higgs potential due to top-stop
  loops, which affect the Higgs masses (for a detailed discussion and references, see e.g.
  \cite{Pierce:1996zz}).
They amount to replacing $m_{H_{u,d}}^2$ in 
Eqns.~\eqref{smallBmu}--\eqref{Bmu} by $\bar m_{H_{u,d}}^2$, where
\be\label{mHbar}
\bar m_{H_u}^2=m_{H_u}^2(M_S)-\frac{t_2}{v\sin\beta}\,,\qquad\bar m_{H_d}^2=m_{H_d}^2(M_S)-\frac{t_1}{v\cos\beta}\,.
\ee
 Here the tadpole terms
$t_i$ are computed from the minimization conditions for the full one-loop 
effective potential \cite{Pierce:1996zz}. 
Using the one-loop results of \cite{Pierce:1996zz}, it turns out that
the dominant corrections to the Higgs masses are
obtained in the limit 
where the top Yukawa coupling is the only non-vanishing coupling, and where
the stop squarks are approximately unmixed and degenerate with mass $m_{\tilde t}=M_S$.
In the $\ol{\rm MS}$ scheme at the renormalization scale $M_S$, one finds
\be\label{tadpoles}
t_1\approx 0\,,\qquad\frac{t_2}{v\sin\beta}\approx \frac{3\,y_t^2}{8\pi^2}m_{\tilde t}^2\,,
\ee
i.e., only $m_{H_u}^2$ is significantly modified by the finite
corrections to the Higgs potential.


In this paper we are interested in the question how electroweak
symmetry breaking can occur at a scale significantly below the
scale of supersymmetry breaking, i.e.~how the conditions $m^2 < 0$ and $|m^2|
\ll M_S^2$ can be realized from Eq.~\eqref{Fermi} at small $|\mu|$.
In particular, how can relation \eqref{smallmHu} be satisfied?
An important observation is that in Eq.~\eqref{mHusq2} the
scalar contributions approximately cancel for equal stop
and $H_u$ masses, $\wh m_{H_u}^2=\wh m_{Q_3}^2=\wh
m_{U_3}^2\equiv m_0^2$. 
This is the basis of the `focus point' idea \cite{Chan:1997bi}. However, as the matching
scale increases, the cancellation between the scalar soft mass
contributions becomes less precise. The actual focussing point of the RG trajectories, 
where the $m_0$ coefficient vanishes, is only obtained for $M_S$ close to the electroweak scale.

\begin{figure}
\centering
 \includegraphics[width=.7\textwidth]{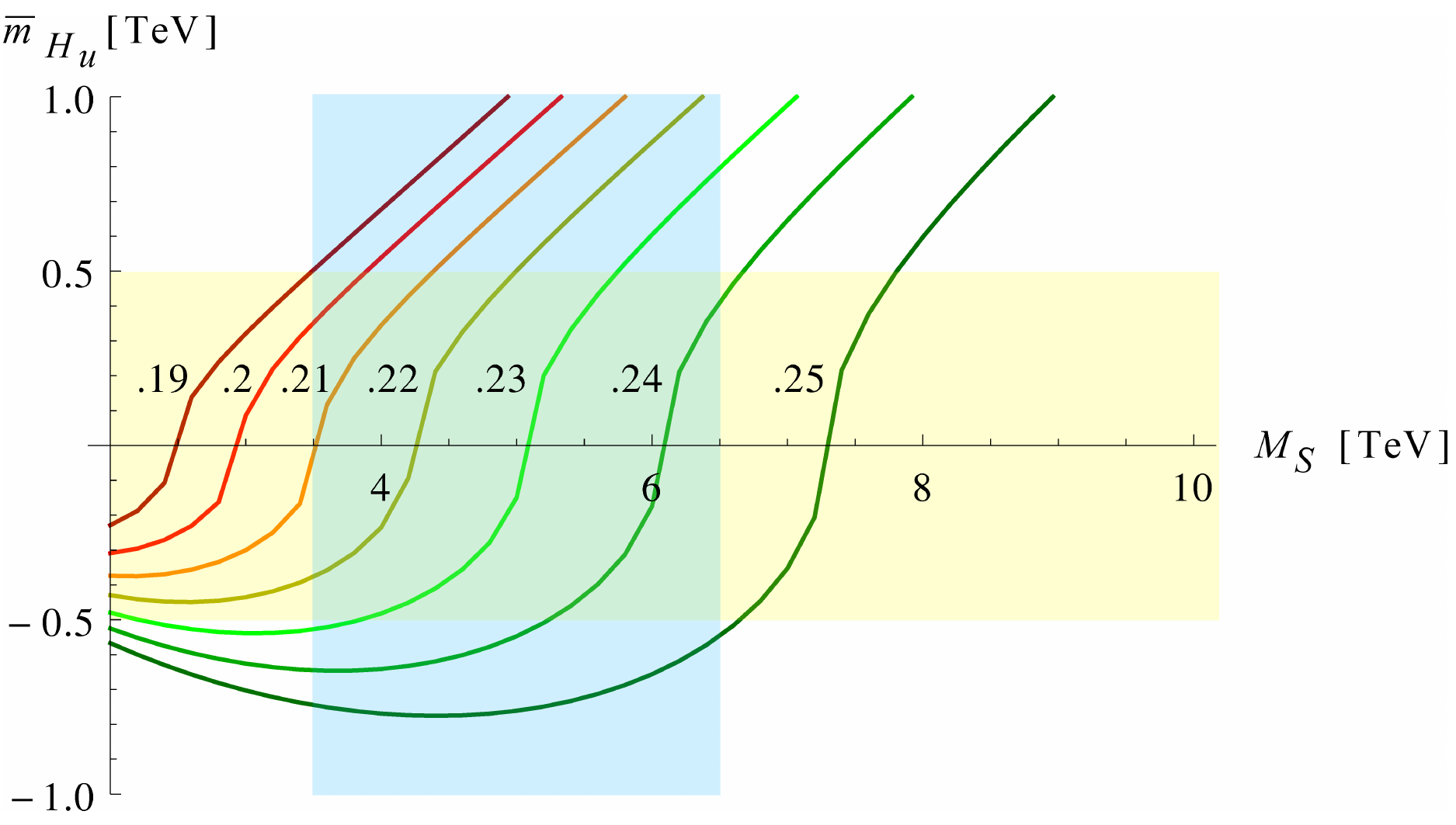}
 \includegraphics[width=.7\textwidth]{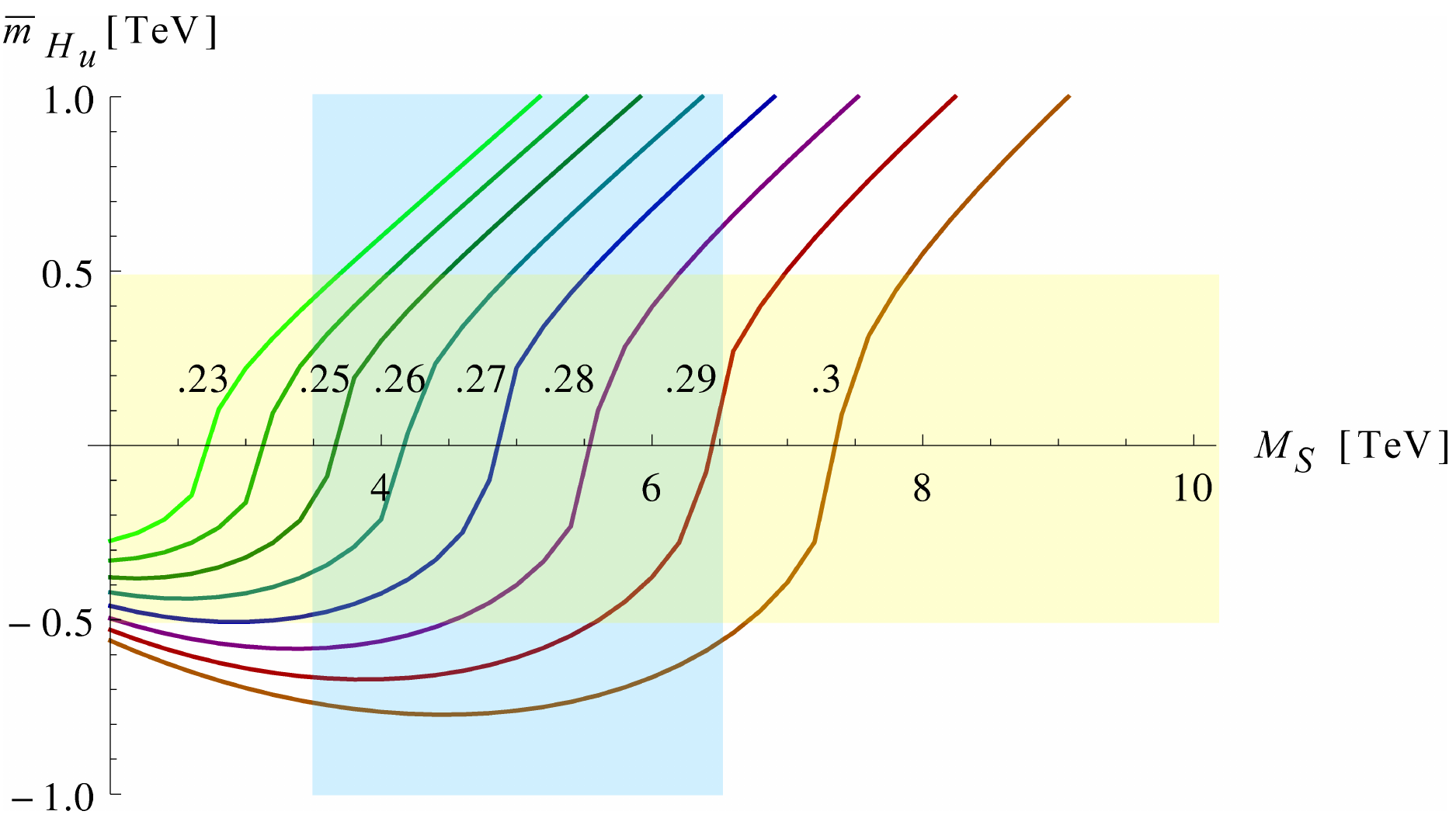}
 \includegraphics[width=.7\textwidth]{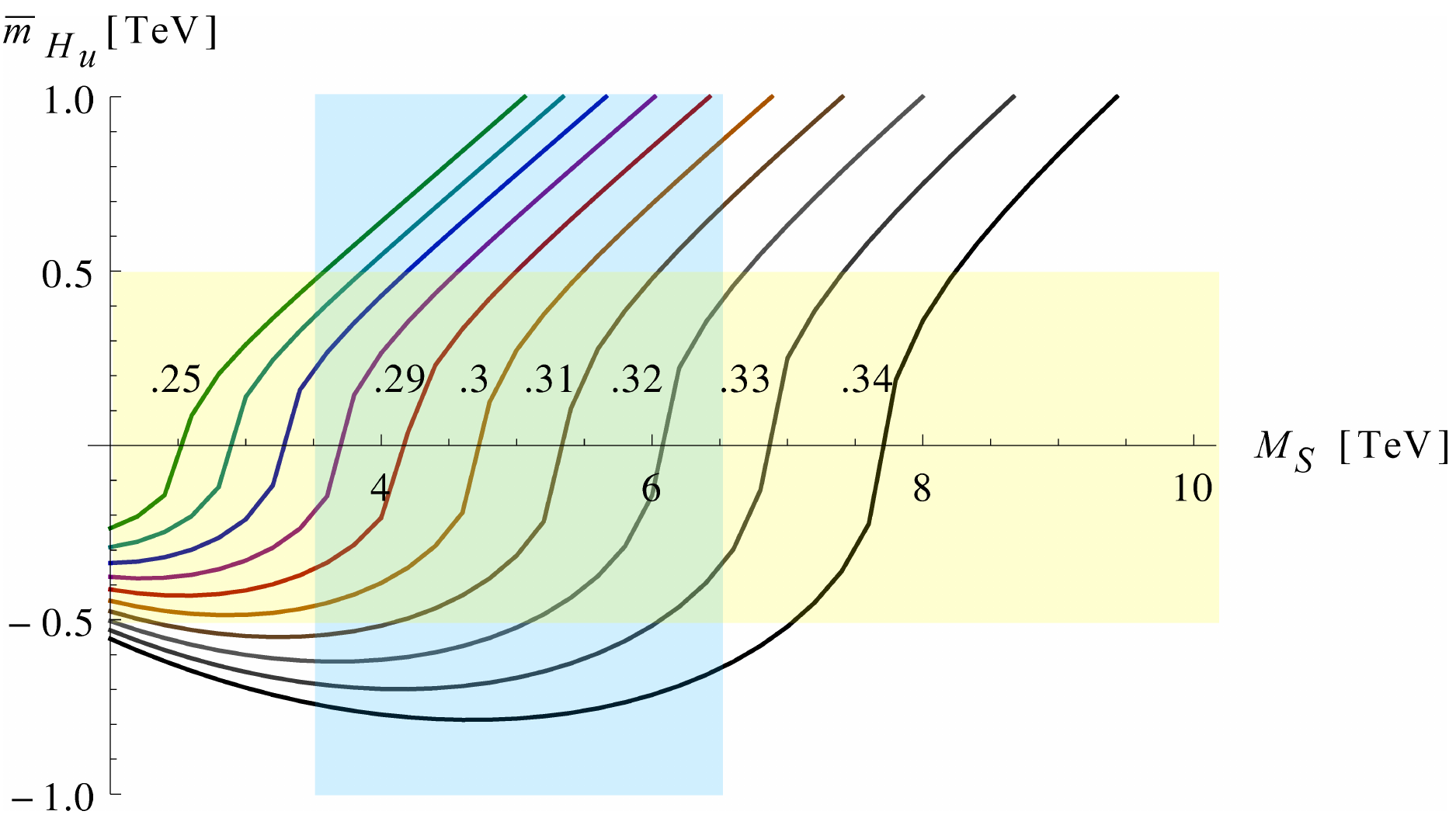}

\caption{$\bar m_{H_u}$ (more precisely $\bar m_{H_u}^2/\sqrt{|\bar m_{H_u}^2|}$) as a function
of the matching scale $M_S$ for various values of the parameter $\kappa=M_{1/2}/m_0$. Top:
 $\wh A_t=-M_{1/2}$, center: $\wh A_t=0$, bottom: $\wh A_t=+M_{1/2}$. Here 
$\tan\beta=15$ and $M_{\rm GUT}=1.5\times 10^{16}$ GeV. We have indicated the 
range of $M_S$ preferred by the Higgs mass (which we took to be $5\pm 1.5$ TeV)
in blue, and a range of $|\bar m_{H_u}|$ around the electroweak scale in yellow.}
\label{mHu_mS}
\end{figure}

As Eq.~\eqref{mHusq2} shows,
the remaining positive contribution to $m_{H_u}^2$ by scalar masses can be compensated by the
negative contribution from gaugino masses. Assuming universal
gaugino masses as suggested by unification, $\wh M_3 = \wh M_2 = \wh M_1 = M_{1/2}$, and taking
the correction Eq.~\eqref{tadpoles} into account, one obtains
for $M_S=5$ TeV
\be\label{unicancel}
\left. \bar m_{H_u}^2\right|_{M_S}= - 1.08\,M_{1/2}^2  + 0.33\,M_{1/2}\wh{A}_t
- 0.12\, {\wh A}_{t}^2 + 0.08\, m_0^2 \ ,
\ee
subject to the uncertainties mentioned above. 
In the following we shall be interested in the case 
$|\wh{A}_t|\lesssim M_{1/2}$. A cancellation between
the gaugino and the scalar contribution then occurs for a
particular ratio $M_{1/2}/m_0$:
\be\label{M12m0}
M_{1/2}=\kappa\, m_0\ ,\qquad\qquad \frac{1}{5}\lesssim\kappa\lesssim \frac{1}{3}\,.
\ee
This can also be seen from Fig.~\ref{mHu_mS}, which shows $\bar m_{H_u}^2(M_S)$ 
as a function of $M_S$ for different values of the ratio $\kappa=M_{1/2}/m_0$
at negative, vanishing, and positive $\wh A_t$.

Models predicting a relation of this type therefore show some promise for 
obtaining a little hierarchy. In Section \ref{6dmodel} we will present an example
with all the required properties: A moderate suppression for the gaugino masses and 
trilinear terms of roughly the correct size, and a good motivation for near-universal 
GUT-scale soft masses of the third generation squarks and Higgs fields. For now
we still need to justify a remaining key assumption, namely that of small $\mu$
and small $B\mu$.

\subsubsection*{Why should $\mu$ and $B\mu$ be small?}

It is well known that the higgsino mass $\mu$ plays a special role among the dimensionful parameters of the MSSM. It preserves supersymmetry, but it breaks a $\U{1}$ Peccei-Quinn (PQ) symmetry under which the Higgs bilinear is charged. The soft masses and trilinear soft terms, by contrast, break supersymmetry but preserve $\U{1}_{\rm PQ}$. The Higgs soft mass mixing parameter $B\mu$ breaks both SUSY and PQ symmetry. 

For concreteness, assume that SUSY is broken by some singlet spurion $X$ with $\vev X=F_X\theta^2$, and that $\U{1}_{\rm PQ}$ is broken supersymmetrically by some spurion $Y$, such that the following terms are allowed in the Lagrangian:
\be\label{PQLag}
{\cal L}=\int d^2\theta\; \frac{Y^p}{M^{p-1}}\left(1 +
  \frac{X}{M}\right) H_u H_d +\int d^4\theta\; \frac{X}{M}\left(|H_u|^2+|H_d|^2\right)\,\hc
\ee
The K\"ahler terms in Eq.~\eqref{PQLag} can be absorbed in the superpotential terms by a field redefinition. The power $p$ depends on the PQ charges of $H_u H_d$ and of $Y$. Bare $\mu$ and $B\mu$ terms $\left.\mu H_u H_d\right|_{\theta^2}$ and $\left.X H_u H_d\right|_{\theta^2}$ are forbidden by $\U{1}_{\rm PQ}$, which also forbids the operators $\left.X^\dag H_u H_d\right|_{\theta^2\bar\theta^2}$ and $\left.|X|^2\, H_u H_d\right|_{\theta^2\bar\theta^2}$. Consequently, the effective $\mu$ parameter is 
\be
\mu\sim\frac{Y^p}{M^{p-1}} \ ,
\ee
and $B\mu$ is proportional to both $\mu$ and the SUSY-breaking vev, 
\be 
B\mu\sim\frac{Y^p}{M^p} F_X \sim \mu M_S \,,
\ee 
where $F_X/M\sim M_S$ is the scale of the scalar and gaugino soft mass parameters.

Choosing $Y$ such that
\be \label{musmall}
 |\mu|^2 \ll M_S^2 
\ee 
 is technically natural, since PQ breaking
is a priori unrelated to SUSY breaking. The `$\mu$ problem' 
is usually formulated as the need for an explanation why the SUSY-breaking 
soft masses are of the same order as $\mu$. Here this is not the 
case: In contrast to the common SUSY model building approach, we obtain 
$\mu$ and the SUSY breaking soft terms from two independent scales. 
As soon as we allow for a little hierarchy, the $\mu$ problem becomes 
less severe as we will argue momentarily. Indeed the most interesting 
parameter choice has $\mu$ maximally separated from $M_S$, to the extent
that is allowed by experimental data.

With the conditions \eqref{smallmHu} and \eqref{musmall}, electroweak 
symmetry can be broken
with all three terms in Eq.~\eqref{Fermi} being of the order of the 
electroweak scale. The required fine-tuning is no worse than the fine-tuning 
needed in the more common case where $\mu$ is of the order of the soft breaking 
terms, and cancelled against a similarly large $\bar m_{H_u}^2$. In our case we are 
instead cancelling large radiative contributions to the $\bar m_{H_u}^2$ parameter 
against each other. 

Remarkably, if the conditions \eqref{musmall} are satisfied with
$\bar m_{H_u}^2$ sufficiently small, then the electroweak scale is parametrically 
given not by $M_S$ but by $\mu$. This is most easily seen by setting $\bar m_{H_u}^2= 0$, 
$m_{H_d}=\eta M_S$, $B\mu=\zeta|\mu| M_S$ at the scale $M_S$, with $\eta$ and $\zeta$ 
of the order one (or at least small compared to $M_S/\mu$ --- in the next section
we will consider a model where $\zeta\sim 1/\kappa$, with $\kappa\approx 0.25$
as in Eq.~\eqref{M12m0}). One then obtains
\be\label{higgsmassmatrix}
m_H^2=\left(\begin{array}{cc}|\mu|^2 & \zeta|\mu| M_S \\ \zeta|\mu|
    M_S & \eta^2 M_S^2\end{array}\right)\ ,
\ee
leading to
\be
- m^2 \simeq \left(\frac{\zeta^2}{\eta^2}-1\right)|\mu|^2\ .
\ee
For $\zeta^2 > \eta^2$
the Higgs mass matrix Eq.~\eqref{higgsmassmatrix} has a negative
eigenvalue even though the diagonal entries are both positive. In
fact, for $\zeta^2 \gg \eta^2$ the electroweak scale is given by a
seesaw-type formula,
\be\label{seesaw}
m^2 \simeq -\frac{(\zeta|\mu| M_S)^2}{\eta^2 M_S^2} =-\frac{\zeta^2}{\eta^2}\,|\mu|^2 < 0 \ .
\ee
A very similar pattern has previously been investigated in the context of
gauge-mediated supersymmetry breaking, where 
the hierarchy between $m_{H_d}$ (or equivalently $M_S$) and $|\mu|$ is not due to a PQ symmetry
but due to a loop factor \cite{Csaki:2008sr}.  Let us
  emphasize that a sufficiently large value of $B\mu$, and therefore
  $\zeta$, is crucial for electroweak symmetry breaking, which takes
  place irrespective of the sign of $m^2_{H_u}$.

As already emphasized we have no symmetry reason for $\bar m_{H_u}^2= 0$. In the more general
case $\left|\bar m_{H_u}^2\right| \ll M_S^2$, electroweak symmetry breaking imposes
a lower bound on $|\mu|^2$,
\be
|\mu|^2 > \frac{\eta^2}{\zeta^2 - \eta^2} \bar m_{H_u}^2 \ .
\ee
Note that there is also a phenomenological lower bound on $|\mu|$:
Since $\tan\beta$ is parametrically given by
$m_{H_d}^2 / B\mu\sim \eta^2 M_S / (\zeta|\mu|)$, and should not exceed a
value $\approx 60$ in order to avoid non-perturbative Yukawa
couplings, the hierarchy between $\mu$ and $M_S$ cannot be 
too large. Thus, for fixed $M_S$, $\mu$ is bounded from below.
The most relevant bound for the model of the next section will
however turn out to be the direct experimental lower limit
$|\mu|\gtrsim 100$ GeV from chargino searches at LEP. 

At this point let us briefly return to the $\mu$ problem. If we 
set $\bar m_{H_u}^2=0$ and ignore the associated fine-tuning for a
moment, it is clear from the Higgs mass matrix Eq.~\eqref{higgsmassmatrix} 
and from Eq.~\eqref{seesaw} that the soft mass scale may be decoupled
from the scale of electroweak symmetry breaking (which is essentially
given by $\mu$). In a hypothetical universe with very light down-type 
quarks, there would also be no restriction on the ratio $M_S/\mu\sim\tan\beta$,
so $M_S$ could in principle be very large, and the $\mu$ problem would
be circumvented. Realistically, however, this line of reasoning 
is invalidated to some extent by the experimentally known bottom and top 
quark masses. The known value of $m_b$ leads to an upper bound on $\tan\beta$, 
while the known value of $m_t$ implies that the top Yukawa coupling is large, 
and that a relation such as $\bar m_{H_u}^2=0$ will therefore be spoiled
by large loop corrections. These two arguments point towards a soft mass 
scale $M_S$ which is not too far above the electroweak scale; the 126
GeV Higgs mass further fixes the `little hierarchy' to amount to 1--2 decades. 
In summary, the $\mu$ problem 
is still present, but somewhat alleviated when allowing for a little 
hierarchy between $M_S$ and the Fermi scale (as seems to be forced upon 
us by LHC data). 

\section{Supersymmetry breaking in higher-dimensional GUTs}\label{6dmodel}

We shall now present an explicit example which realizes
the conditions for a seesaw-type pattern of electroweak symmetry
breaking discussed in the previous section.
Consider a six-dimensional (6d) GUT model, with the third
quark-lepton generation and the Higgs fields located in the bulk and the first
two families localized at 4d branes or orbifold fixed points.
Such a model has been derived as an intermediate step
\cite{Buchmuller:2007qf} in a compactification of the heterotic
string to the supersymmetric standard model in four dimensions 
\cite{Buchmuller:2005jr,Lebedev:2006kn}. Supersymmetry 
is supposed to be broken by the $F$-term of a chiral
superfield located at some fixed point. 

In the following we shall restrict our discussion to the case of strong coupling at
the cutoff scale. The couplings of the supersymmetry breaking brane field to Higgs,
matter and gauge fields can then be estimated by means of `naive
dimensional analysis' (NDA) following \cite{Chacko:1999mi}.
The localization of the
fields fixes the structure of the Lagrangian
\begin{equation} \label{eq:L6d}
\mathcal{L}_{6d}= \
\mathcal{L}_\mathrm{bulk}\left(W_\alpha,\Phi\right) +
\sum_i \delta^{2}(y-y_i) \
\mathcal{L}_i\left(W_\alpha,\Phi,\phi\right) \ ,
\end{equation}
where $y_i$ are the positions of the 4d branes, and $W_\alpha$, $\Phi$ and
$\phi$ denote bulk gauge fields, bulk chiral fields, and brane chiral fields,
respectively. Matching 6d and 4d theories at the compactification
scale, the gauge couplings and Planck masses are related by
\begin{equation}
\frac{1}{g_6^2} V_{2} = \frac{1}{g_4^2} \ , \quad M_6^4 V_{2} = M_4^2 \ ,
\end{equation}
where $V_{2}$ is the volume of the two compact dimensions.\footnote{In
 the considered GUT model one has $V_{2} = 2\pi^2 R_5R_6$, where
 $R_5$ and $R_6$ are the radii of the orbifold. The model has a Wilson
 line in the direction of $R_6$ which breaks the GUT symmetry. With 
 $R_5 \geq R_6$, the mass of the
 lowest lying Kaluza-Klein state is $1/(2R_5)$. Identifying this mass
 with the GUT scale $M_\mathrm{GUT} \simeq 1\times 10^{16}\mathrm{GeV}$,
 one obtains $V_{2}^{-1/2} \simeq 5\times 10^{15}\mathrm{GeV}$ (see \cite{Hosteins:2009xk}).} 

In order to define the theory one has to introduce a UV cutoff
$\Lambda$. If loop corrections at the scale $\Lambda$ are suppressed by
$\epsilon$, the Lagrangian Eq.~\eqref{eq:L6d} can be expressed in terms of
dimensionless fields $\wh W_\alpha/\Lambda^{3/2}$, $\wh\Phi/\Lambda$ and $\wh\phi/\Lambda$,
\begin{align}
W_\alpha(x,y) &= \frac{\Lambda^{5/2}}{\left(\epsilon\ell_6\right)^{1/2}}
\frac{\wh W_\alpha(x,y)}{\Lambda^{3/2}}\,, \quad
\Phi(x,y) = \left(\frac{\Lambda^{4}}{\epsilon\ell_6}\right)^{1/2}
\frac{\wh\Phi(x,y)}{\Lambda} \ , \label{scalingbulk} \\
\phi_i(x) &= \left(\frac{\Lambda^2}{\epsilon\ell_4}\right)^{1/2} 
\frac{\wh\phi_i(x)}{\Lambda} \ .\label{scalingbrane}
\end{align}
The fields $W_\alpha$, $\Phi$ and $\phi$ are assumed to have canonical
kinetic terms in 6d and 4d, respectively, and the rescaled fields 
$\wh W_\alpha$, $\wh \Phi$ and $\wh\phi$ have canonical dimensions in
4d. According to NDA the Lagrangian \eqref{eq:L6d} now takes the form
\begin{equation} \label{eq:Ldimless}
\mathcal{L}_{6d} = 
\frac{\Lambda^6}{\epsilon\ell_6} \
\mathcal{\wh L}_\mathrm{bulk}\left(\frac{\wh W_\alpha}{\Lambda^{3/2}},\frac{\wh\Phi}{\Lambda},
\frac{\partial}{\Lambda}\right) +
\sum_i \delta^{2}(y-y_i) \ \frac{\Lambda^4}{\epsilon\ell_4} \
\mathcal{\wh L}_i\left(\frac{\wh W_\alpha}{\Lambda^{3/2}},\frac{\wh\Phi}{\Lambda},\frac{\wh\phi}
{\Lambda},\frac{\partial}{\Lambda}\right) \ ,
\end{equation}
where all couplings are $\mathcal{O}(1)$ and
$\ell_D = 2^D \pi^{D/2} \, \Gamma(D/2)$ is a geometrical loop factor, with
\begin{equation}
\ell_6 = 128 \pi^3 \ , \qquad \ell_4 = 16 \pi^2 \ .
\end{equation}
Strong coupling at the cutoff scale $\Lambda$ corresponds to $\epsilon \simeq 1$. 

In our 6d GUT model, the couplings of the SUSY
breaking brane field $X$ to the bulk fields are given by\footnote{This
 model has two pairs of equivalent fixed points \cite{Buchmuller:2007qf}. Hence,
 there will be at least two SUSY breaking fields, at a pair of
 equivalent fixed points.  For the following discussion this complication is irrelevant
and will be ignored.}
\begin{align} \label{LGUTdimless}
-\mathcal{L}_{sb} = 
\frac{\Lambda^4}{\epsilon\ell_4}  \bigg\{
&\int \frac{d^2\theta}{\Lambda} \left(\frac{\wh\mu}{\Lambda} \frac{\wh{H}_u}{\Lambda} 
\frac{\wh{H}_d}{\Lambda}\left(1 + \frac{\wh{X}}{\Lambda}\right)
+ \frac{\wh{X}}{\Lambda}\left(\tr{\left[\frac{\wh{W}^\alpha}{\Lambda^{3/2}} 
\frac{\wh{W}_\alpha}{\Lambda^{3/2}}\right]} 
+ \frac{\wh{Q}_3}{\Lambda}\frac{\wh{H}_u}{\Lambda}\frac{\wh{U}_3}{\Lambda}
\right)  \hc \right) \nonumber\\
 + &\int \frac{d^4\theta}{\Lambda^2}\ \frac{|\wh{X}|^2}{\Lambda^2}
\left(\frac{|\wh{H}_u|^2}{\Lambda^2}+\frac{|\wh H_d|^2}{\Lambda^2}
+\frac{|\wh{Q}_3|^2}{\Lambda^2}+\frac{|\wh{U}_3|^2}{\Lambda^2}+\frac{|\wh
 {D}_3|^2}{\Lambda^2}+
\frac{|\wh{L}_3|^2}{\Lambda^2}+\frac{|\wh{E}_3|^2}{\Lambda^2}\right)
\bigg\} \ ,
\end{align}
where $H_u$, $H_d$, $W_\alpha$, $Q_3$, $U_3$, $D_3$, $L_3$ and $E_3$ denote
Higgs fields, gauge fields and third generation quark and lepton
fields, respectively.  $H_u$ is part of the 6d gauge multiplet, 
$Q_3$ and $U_3$ belong to the same hypermultiplet, and
the cubic term $Q_3H_uU_3$ is part of the 6d gauge
interactions.\footnote{Trilinear terms for the other matter fields are 
also allowed but will not be written explicitly.}
From the gauge kinetic term one reads off the gauge
coupling
\begin{align}
g_6 \sim \frac{\left(\epsilon\ell_6\right)^{1/2}}{\Lambda} \ .
\end{align}
The mass parameter $\wh\mu$ is an additional free parameter which can
be much smaller than the cutoff scale $\Lambda$ due to an accidental
PQ symmetry as discussed in Section \ref{littleh}.
From Eqs.~\eqref{scalingbulk} and \eqref{LGUTdimless} one obtains the
Lagrangian for canonically normalized bulk fields,
\begin{align} \label{LGUTcan}
-\mathcal{L}_{sb} = 
\frac{\ell_6}{\ell_4}\frac{1}{\Lambda^{D-4}}\bigg\{
&\int d^2\theta \left(\hat{\mu} H_u H_d \left(1 
+ \frac{\wh{X}}{\Lambda}\right) + \frac{\wh{X}}{\Lambda}
\left(\tr{\left[W^\alpha W_\alpha\right]} 
+ g_6 Q_3 H_u U_3 \hc \right)  \right) \nonumber\\
 + &\int d^4\theta\frac{|\wh{X}|^2}{\Lambda^2}
\left(|H_u|^2 + |H_d|^2 + |Q_3|^2 + |U_3|^2 + |D_3|^2 +
|L_3|^2 + |E_3|^2 \right) \bigg\} \ .
\end{align}
Finally, the replacement
$\Phi(x,y) \rightarrow V_2^{-1/2} \Phi(x)$ yields the couplings of
canonically normalized zero modes,
\begin{align} \label{LGUTzero}
-\mathcal{L}_{sb} = 
\frac{\ell_6}{\ell_4}\frac{1}{\Lambda^{2}V_2} \bigg\{
&\int d^2\theta \left(\hat{\mu} H_u H_d \left(1 
+ \frac{\wh{X}}{\Lambda}\right) + \frac{\wh{X}}{\Lambda}
\left(\tr{\left[W^\alpha W_\alpha\right]} 
+ g_4 Q_3 H_u U_3 \hc \right)  \right) \nonumber\\
 + &\int d^4\theta \frac{|\wh{X}|^2}{\Lambda^2}
\left(|H_u|^2 + |H_d|^2 + |Q_3|^2 + |U_3|^2 + |D_3|^2 +
|L_3|^2 + |E_3|^2 \right) \bigg\} \ .
\end{align}

In Eq.~\eqref{LGUTdimless} we have assumed a universal coupling of the
SUSY breaking field to bulk fields. The focus point cancellation
discussed in Section~2 requires approximately equal mass terms of $H_u$, $Q_3$
and $U_3$ at a level of about 5\%. In the considered model the
equality of mass terms is  guaranteed by a symmetry only
  for $U_3$ and $E_3$, which belong to the same $\mathrm{SU(6)}$
  hypermultiplet in six dimensions. For all other fields a  dynamical
  reason is needed.
The couplings of brane
and bulk fields depend on the profile that the bulk fields aquire in
connection with the stabilization of the compact dimensions. These
profiles depend on the presence of localized Fayet-Iliopoulos (FI) terms
\cite{Lee:2003mc}. Such FI-terms are also crucial to reconcile the tree-level
gauge-top unification of the model, $y_t = g_4$ (cf.~Eq.~\eqref{LGUTzero}), with the large values
of $\tan\beta$ considered in Section~2 \cite{Hosteins:2009xk}.
 It is conceivable that the FI terms present in the model \cite{Buchmuller:2007qf}
lead to approximately equal mass terms, 
but a detailed study of the compactification dynamics is beyond
  the scope of this paper.

Replacing now the brane field $X$ by its SUSY breaking vacuum expectation value $F_X$, 
we obtain from Eq.~\eqref{LGUTzero} the wanted mass parameters of the
zero modes for gaugino fields, Higgs and higgsino fields and third
generation scalar quark-lepton fields,
\begin{align}
\mathcal{L}_{soft} = \ & -\left(\frac{1}{2}\mu\, h_u h_d + B\mu\, H_u H_d +\frac{1}{2} M_{1/2} \tr \lambda^a \lambda^a  
+ A_t y_t\, Q_3 H_u U_3 \hc\right)\nonumber \\
&- \left(m^2_0 + |\mu|^2\right) \left(|H_u|^2 + |H_d|^2\right) 
- m^2_0 \left(|Q_3|^2 + |U_3|^2 + |D_3|^2 + |L_3|^2 + |E_3|^2 \right)  \label{musoft}\ ,
\end{align}
where $y_t = g_4$, and
\begin{align}
m_0^2 &\sim \kappa^2 \left(\frac{\wh{F}_X}{\Lambda}\right)^2 \ , \qquad
M_{1/2} \sim \kappa^2\ \frac{\wh{F}_X}{\Lambda} \sim \kappa\ m_0
\ , \qquad A_t \sim M_{1/2}  \label{scalarsoft} \\
\mu &\sim \kappa^2\ \hat{\mu} \ , \qquad
B\mu \sim \kappa^2\ \hat{\mu} \frac{\wh{F}_X}{\Lambda}
\sim \frac{1}{\kappa}\ \mu m_0 \ ,
\end{align}
with
\begin{align}
\kappa^2 = \frac{\ell_6}{\ell_4}\frac{1}{\Lambda^{2}V_{2}}  
= \frac{\ell_6}{\ell_4} \left(\frac{M_6}{\Lambda}\right)^2\frac{1}{M_4 V_2^{1/2}} \ .
\end{align}
For $\Lambda = M_6$ and  GUT-scale extra dimensions, i.e. $V_2^{-1/2} \simeq
5\times 10^{15}\,\mathrm{GeV}$, this yields $\kappa^2 \simeq 0.06$. For
the particularly interesting gaugino-scalar mass relation we then
obtain
\begin{equation}
M_{1/2} \sim\ 0.25\ m_0\ .
\end{equation}  
Let us emphasize that this relation is not at all generic, but based
on a 6d GUT picture, supersymmetry breaking by a brane field and the assumption
of strong coupling at the UV cutoff which is chosen to be the 6d Planck mass.
The prediction for analogous models with a different number of
GUT-scale extra dimensions is not too different, however: the general expression
for $D$ dimensions and $\Lambda=M_D$ reads
\be
\kappa^2=\frac{\ell_D}{\ell_4} \left(\frac{1}{M_4^{D-4} V_{D-4}}\right)^{\frac{2}{D-2}}
\ee
and yields $\kappa^2=0.08$ ($0.09$) for $D=5$ ($D=7$), assuming the same 
compactification radius as above. For even larger $D$ the loop factor
enhancement becomes dominant, and $\kappa^2$ grows rather large.
The precise choice of the compactification scale 
sensitively affects the prediction for $\kappa^2$, and its proper value is 
dependent on the model details. The following discussion applies to our $D=6$
orbifold model with $V_2^{-1/2} \simeq 5\times 10^{15}\,\mathrm{GeV}$.

Comparing with the electroweak symmetry breaking pattern of the last chapter,
we find that in this model the $\kappa$ parameter is of the correct order
of magnitude to explain a small $\bar m_{H_u}^2$ and a little hierarchy
between the electroweak scale and the soft mass scale. The $\zeta$ parameter
is somewhat large at $\zeta\sim 1/\kappa \sim 4$; therefore if
 $\bar m_{H_u}^2$ were completely negligible, we would obtain a slightly too
large electroweak scale from Eq.~\eqref{seesaw}:
\be
v^2=-\frac{m^2}{\lambda}\approx \frac{\left(4\mu\right)^2}{\lambda}
\ee
which is not compatible with the experimental lower bound $|\mu|\gtrsim 100$ 
GeV. However, a finite negative $\bar m_{H_u}^2$ can easily cure this.

An important quantity is the gravitino mass. From
Eq.~\eqref{scalarsoft} one obtains for the scalar mass parameter
\begin{align}
m_0 \simeq \sqrt \ell_6 \left(\frac{M_6}{\Lambda}\right)^2
\frac{F_X}{M_4} \ .
\end{align}
Together with $m_{3/2} = F_X/(\sqrt 3 M_4)$ this yields
\begin{align}\label{gravmass}
m_{3/2} \simeq \frac{1}{\sqrt{3\ell_6}} \left(\frac{\Lambda}{M_6}\right)^2 m_0
\simeq 0.01\ \left(\frac{\Lambda}{M_6}\right)^2 m_0 \ .
\end{align}
Hence, unless the cutoff significantly exceeds the 6d Planck mass, the
gravitino will be the lightest superparticle. The result
\eqref{gravmass} is consistent with the analysis carried out in
\cite{Buchmuller:2005rt}.\footnote{Note that the predictions of masses
obtained from naive dimensional analysis have an uncertainty of
$\mathcal{O}(1)$. This includes the effect of a colour factor which
was included in the calculations in \cite{Buchmuller:2005rt} and which
has been omitted in the present discussion for simplicity.}

Let us finally consider the first and second quark-lepton generations,
which are localized at two equivalent fixed points (see
\cite{Buchmuller:2007qf}) that may or may not coincide with the
localization of the supersymmetry breaking fields. In the second case
one has
\begin{align}
-\mathcal{L}'_{soft} =  
 \tilde{m}^{2}_0 \sum_{i=1,2} \left(|Q_i|^2 + |U_i|^2 + |D_i|^2 +
|L_i|^2 + |E_i|^2 \right)  \label{12soft}\ ,
\end{align}
with $\tilde{m}_0 = 0$, correponding to the boudary conditions of
gaugino mediation.\footnote{Strictly speaking this is not possible with the exact particle
content of \cite{Buchmuller:2007qf}, since in this model there is no suitable 
brane-localized singlet which could play the role of $X$. However a slight
variation of this model might well contain a suitable candidate.}  
In the first case, the scalar mass terms are
obtained from Eq.~\eqref{LGUTdimless},
\begin{align} 
-\mathcal{L}'_{sb} = 
\frac{\Lambda^4}{\epsilon\ell_4}  \sum_{i=1,2}
 \int \frac{d^4\theta}{\Lambda^2}\ \frac{|\wh{X}|^2}{\Lambda^2}
\left(\frac{|\wh Q_i|^2}{\Lambda^2}+\frac{|\wh U_i|^2}{\Lambda^2}+\frac{|\wh
  D_i|^2}{\Lambda^2}+
\frac{|\wh L_i|^2}{\Lambda^2}+\frac{|\wh E_i|^2}{\Lambda^2}\right) \ .
\end{align}
Performing the transition to canonically normalized fields using
\eqref{scalingbrane}, one finds in the case of strong coupling
($\epsilon = 1$),
\begin{align}\label{heavy}
\tilde{m}_0 \simeq \frac{1}{\kappa} m_0 \sim 4 m_0 \ .
\end{align}
Hence, in this case, unlike gaugino mediation, first- and
second-generation scalars will be heavier than third-generation scalars.

\section{Prospects for phenomenology and outlook}

\begin{figure}
\centering
 \includegraphics[width=.7\textwidth]{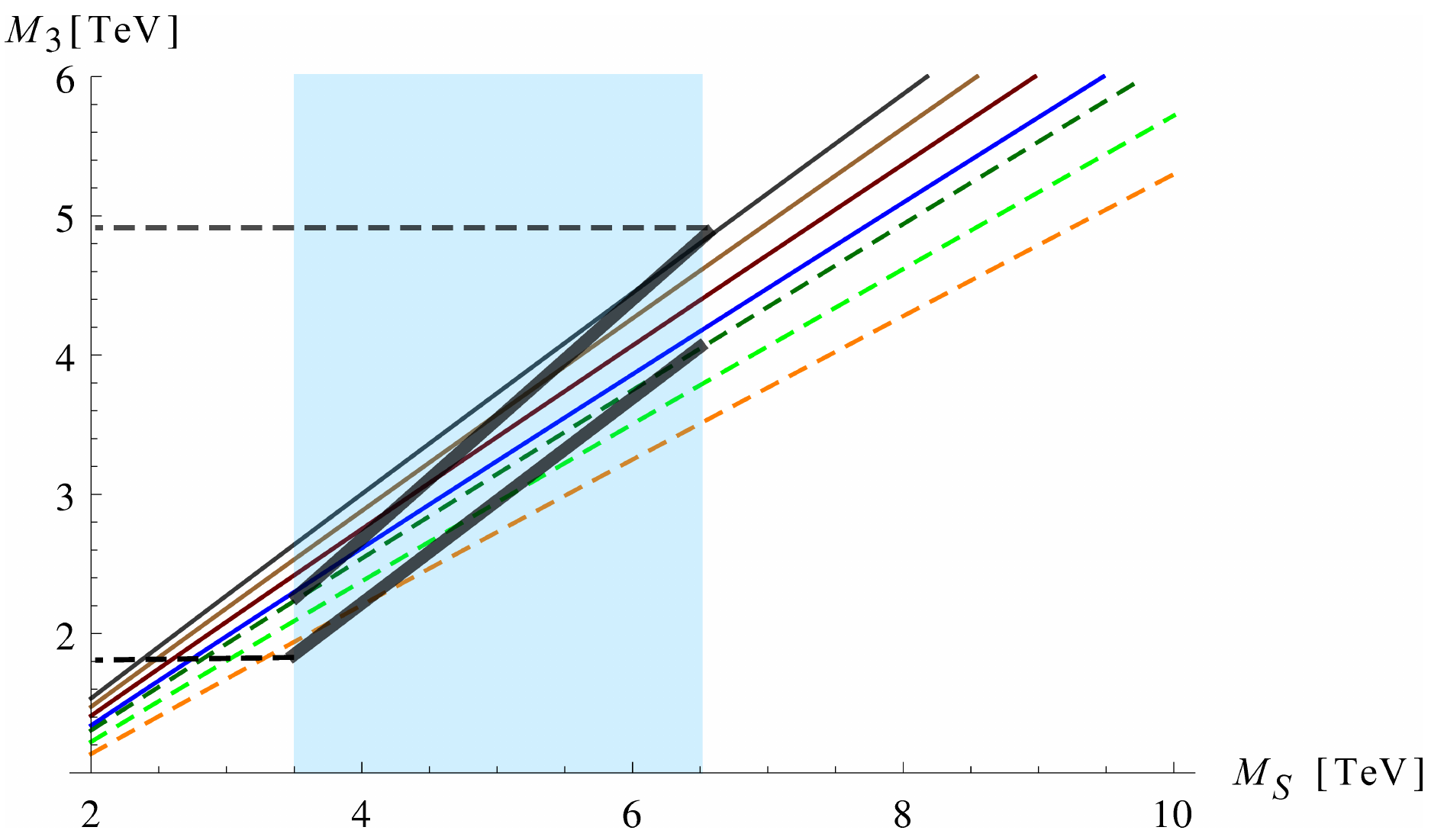}
\caption{The running gluino mass $\left.M_3\right|_{M_S}$ as a function
of the matching scale $M_S$ for various values of the parameter $\kappa=M_{1/2}/m_0$. 
Top to bottom curves: $\kappa$ between $0.33$ and $0.19$ in steps of $0.02$.
The solid lines correspond to $\wh A_t=+M_{1/2}$ and the dashed lines to 
$\wh A_t=-M_{1/2}$, with the colour coding the same as in Fig.~\ref{mHu_mS}. Note that the 
relation between $M_S$ and $M_3$ for fixed $\kappa$ is only approximately 
linear. As before, $\tan\beta=15$, $M_{\rm GUT}=1.5\times 10^{16}$ GeV. We have 
indicated the range of $M_S$ preferred by the Higgs mass (which we took to be 
$5\pm 1.5$ TeV). We have also indicated the predictions for $M_3$ as a function
of $M_S$, for the two cases $\wh A_t=\pm M_{1/2}$ (black strips), and the
minimal and maximal $M_3$ which can be obtained (dashed horizontal lines), when
restricting $|m_{H_u}|$ to be of the order of the electroweak scale
as in Fig.~\ref{mHu_mS}. One finds $1.8$ TeV $\lesssim M_3\lesssim 4.9$ TeV.}\label{M3_mS}
\end{figure}

In the six-dimensional GUT model which we discussed in the previous section,
the localization of fields and the breaking of supersymmetry by a brane field 
determine the pattern of scalar and gaugino masses. The 
Higgs bosons, third generation squarks and sleptons, and gauginos
are bulk fields. Their masses depend on $\kappa=M_{1/2}/m_0$,
as determined by the matching scale $M_S$, the sign of $\wh
A_t$ and $m_0$ (which in turn is related to the matching scale by
renormalization group running). In Fig.~\ref {M3_mS} the gluino mass at $M_S$ is shown as function of
$M_S$ for both signs of $\wh A_t$. The resulting predicted range of gluino masses,
\begin{align}
2~\mathrm{TeV} \lesssim M_3|_{M_S} \lesssim 5~\mathrm{TeV} \ ,
\end{align}
is a consequence of the allowed range of matching scales and the sign
ambiguity of $\wh A_t$. 

The first two generations are brane fields. Their masses strongly
depend on the localization of the supersymmetry breaking field $X$.
There are two possibilities: 
\begin{enumerate}
\item[(A)] The matter fields and $X$ are localized on different
  branes. This implies the
familiar pattern of gaugino mediation, and squarks and sleptons of
the first two generations are lighter than those of the third generation.\footnote{Note that for the same localization of fields, but
 a different mechanism of supersymmetry breaking, third generation squarks and
  sleptons can also be lighter than those of the first two generation \cite{Krippendorf:2012ir}.}
\item[(B)] The matter fields and $X$ are localized on the same brane. According to
  Eq.~\eqref{heavy}, derived in the previous section, the squarks and
  sleptons of the first two generations are then heavier than those of
  the third generation.
\end{enumerate}

A further important parameter is the higgsino 
mass $\mu$. If $\mu$ is generated independently of supersymmetry breaking, generically
one would expect $\mu\sim M_S/\tan\beta$, as discussed in Section \ref{littleh}. In
the model of Section \ref{6dmodel}, since $B\mu$ is enhanced by a factor $1/\kappa$,
we estimate $\mu\sim \kappa M_S/\tan\beta$, which implies that for $M_S=5\pm 1.5$ TeV
and moderately large $\tan\beta$, the $\mu$ parameter should actually be close to the 
electroweak scale ($|\mu|\lesssim 100$ GeV being excluded by chargino searches). A soft 
upper bound can be estimated by conservatively setting $\tan\beta=5$, $\kappa=1/3$ and 
$M_S=6.5$ TeV, which yields $\mu\lesssim 450$ GeV. 

In summary, the mass spectrum we predict is characterized by heavy third-generation
squarks and sleptons, heavy extra Higgs bosons, gluino masses starting from
about 2 TeV, higgsino-like charginos and neutralinos with electroweak-scale masses, 
and squarks and sleptons which are either extremely heavy (B) or generated by
gaugino mediation (A). In the
latter case, standard SUSY searches for jets and missing energy, as well as searches for
direct slepton production, will be promising channels at LHC-14. In any case, the light
higgsinos can be searched for and measured at a linear collider 
\cite{Baer:2011ec,Baer:2012uy}.

\begin{table}[t]
 \centering
\begin{tabular}{ccccc}
\hline
& \multicolumn{3}{c}{light 1st \& 2nd generation} & heavy 1st \& 2nd generation\\ 
\hline
 & $M_S=3.5$ TeV & $M_S=5$ TeV & $M_S=6.5$ TeV & $M_S=5$ TeV\\
 \hline
 $\chi^0_1$ & $127$ & $109$ & $141$ & $185$\\
 $\chi^0_2$ & $140$ & $116$ & $146$ & $189$\\
 $\chi^\pm_1$ & $133$ & $112$ & $144$ & $187$\\
 $\chi^0_3$ & $430$ & $700$ & $990$ & $1100$\\
 $\chi^0_4, \chi^\pm_2$ & $820$ & $1300$ & $1900$ & $2100$\\ 
 $H_0$, $A_0$, $H^\pm$ & $4200$ & $5900$ & $7500$ & $7200$\\
 $\tilde g$ & $2200$ & $3500$ & $4800$ & $5600$\\
 $\tilde u_i$, $\tilde d_i$, $\tilde c_i$, $\tilde s_i$ & $1800$ -- $2000$ & $2800$ -- $3000$ & $3900$ -- $4100$ & $3\times 10^4$\\
 $\tilde t_1$ & $3100$ & $4500$ & $5800$ & $4400$\\
 $\tilde t_2$ & $4000$ & $5600$ & $7300$ & $5900$\\
 $\tilde b_1$ & $4000$ & $5700$ & $7400$ & $6000$\\
 $\tilde b_2$ & $4600$ & $6500$ & $8400$ & $7400$\\
 $\tilde\mu_1$, $\tilde e_1$ & $350$ & $560$ & $800$ & $3\times 10^4$\\
 $\tilde\mu_2$, $\tilde e_2$ & $610$ & $1000$ & $1400$ & $3\times 10^4$\\
 $\tilde \tau_1$ & $4300$ & $5900$ & $7500$ & $7400$\\ 
 $\tilde \tau_2$ & $4400$ & $6000$ & $7700$ & $7500$\\
\end{tabular}
\caption{Example mass spectra computed with {\tt SOFTSUSY 3.3.10} \cite{Allanach:2001kg} for different matching scales. The parameters are $\tan\beta=15$, $\wh A_{t,b}=0$, $M_{1/2}=(1,\,1.6,\,2.25,\,2.45)$ TeV, $m_0=(4.35,\,6.0,\,7.6,\,7.7)$ TeV for the third generation and the Higgs fields. For the three columns on the left, the GUT-scale scalar soft masses of the first two generations vanish, whereas for the rightmost column they are $\tilde m_{0}=30$ TeV. All masses in the table are in units of GeV.
}\label{benchmark}
\end{table}

Table \ref{benchmark} shows a number of superpartner mass spectra. The first three 
columns correspond to three different values of $M_S$ in scenario (A). For a relatively
low matching scale $M_S=3.5$ TeV, gluinos and squarks should be found during the
next LHC run, and sleptons should also be easy to see as the slepton masses are 
already at the border of the present exclusion bounds \cite{Atlas-sleptons,CMS-sleptons}. 
The case of an intermediate
matching scale $M_S=5$ TeV is more challenging, but squarks and gluinos may still
be accessible at high integrated luminosities. The third case of $M_S=6.5$ TeV
places squarks and gluinos out of LHC reach. 

The last column of Table \ref{benchmark} shows a spectrum for the case that the 
first- and second-generation scalar masses are non-vanishing at the GUT scale and given by 
$M_{\rm soft}=30$ TeV (scenario (B)  above). In this case the overall soft mass scale also 
for the third generation and the gluinos is higher. The reason is that we are keeping 
$M_S=5$ TeV fixed, and the first two squark generations significantly decrease the stop 
masses when running down from the GUT scale through two-loop effects, up to a point where
the stop mixing contribution to the lightest Higgs mass can become very significant
 \cite{ArkaniHamed:1997ab, Baer:2012uy}. This case is not covered by our
semi-analytic discussion in Section \ref{littleh}, which does not account for possible large
contributions to the running from the first two generations, but can nevertheless be
dealt with numerically. As is evident from Table \ref{benchmark},
all states are too heavy to be seen at colliders in the foreseeable future, with the possible 
exception of the higgsinos.

Finally, the matching scale $M_S$ also determines the gravitino mass.
From Eq.~\eqref{gravmass} one obtains
\begin{align}
40~\mathrm{GeV} \simeq m_{3/2} \simeq 80~\mathrm{GeV} \ .
\end{align}
Here we have chosen the 6d Planck mass as the cutoff scale, and we have
varied $m_0$ between $4$ TeV and $8$ TeV according to Table
\ref{benchmark}.

The starting point of our discussion has been the compatibility of the
measured Higgs boson mass, and the associated large matching scale
$M_S$, with a Fermi scale significantly smaller than $M_S$. 
We have shown that for a small
higgsino mass $\mu$, not controlled by supersymmetry breaking, and
universal Higgs and stop masses at the GUT scale, a
small Fermi scale arises for suitable relations between gaugino and
scalar masses. It is interesting that a simple example
can be obtained within the context of a higher-dimensional GUT model. 
The matching scale, together with the related gaugino-scalar mass ratio,
and the value of $\tan\beta$ determine the superparticle mass spectrum.
If the matching scale turns out to be lower than about $5$ TeV, this
scenario will be probed by the upcoming next LHC run, with searches 
for gluinos, squarks and also sleptons being promising channels. Moreover,
our setup favours light higgsinos, which can be searched for at a linear collider.
The lightest superparticle is the gravitino.\\

\medskip

\subsection*{Acknowledgements}

We thank J.~Kersten, M.~Ratz und M.~Winkler for helpful comments. This work 
was supported in part by the German Science Foundation (DFG) within the 
Collaborative Research Center 676 ``Particles, Strings and the Early Universe''.
The work of FB was supported in part by ERC Advanced Grant 267985 
``Electroweak Symmetry Breaking, Flavour and Dark Matter''.




\begin{thebibliography}{99}

\bibitem{Dimopoulos:1981zb}
  S.~Dimopoulos and H.~Georgi,
  Nucl.\ Phys.\ B {\bf 193} (1981) 150.

\bibitem{Ibanez:1981yh}
  L.~E.~Ib\'a\~nez and G.~G.~Ross,
  Phys.\ Lett.\ B {\bf 105} (1981) 439.

\bibitem{Dimopoulos:1981yj}
  S.~Dimopoulos, S.~Raby and F.~Wilczek,
  Phys.\ Rev.\ D {\bf 24} (1981) 1681.



\bibitem{Aad:2013wta}
  G.~Aad {\it et al.}  [ATLAS Collaboration],
  arXiv:1308.1841 [hep-ex],
ATLAS-CONF-2013-089, 
ATLAS-CONF-2013-062, 
ATLAS-CONF-2013-047.  





\bibitem{Chatrchyan:2013lya}
  S.~Chatrchyan {\it et al.}  [CMS Collaboration],
  Eur.\ Phys.\ J.\ C {\bf 73} (2013) 2568
  [arXiv:1303.2985 [hep-ex]],
CMS-PAS-SUS-13-013,  
CMS-PAS-SUS-13-012,  
CMS-PAS-SUS-13-004.  


\bibitem{Aad:2012tfa}
  G.~Aad {\it et al.}  [ATLAS Collaboration],
  Phys.\ Lett.\ B {\bf 716} (2012) 1
  [arXiv:1207.7214 [hep-ex]].


\bibitem{Chatrchyan:2012ufa}
  S.~Chatrchyan {\it et al.}  [CMS Collaboration],
  Phys.\ Lett.\ B {\bf 716} (2012) 30
  [arXiv:1207.7235 [hep-ex]],
  JHEP {\bf 1306} (2013) 081
  [arXiv:1303.4571 [hep-ex]].

\bibitem{ArkaniHamed:2004fb}
  N.~Arkani-Hamed and S.~Dimopoulos,
  JHEP {\bf 0506} (2005) 073
  [hep-th/0405159];
  G.~F.~Giudice and A.~Romanino,
  Nucl.\ Phys.\ B {\bf 699} (2004) 65
   [Erratum-ibid.\ B {\bf 706} (2005) 65]
  [hep-ph/0406088].


\bibitem{Chan:1997bi}
  K.~L.~Chan, U.~Chattopadhyay and P.~Nath,
  Phys.\ Rev.\ D {\bf 58} (1998) 096004
  [hep-ph/9710473];
  J.~L.~Feng, K.~T.~Matchev and T.~Moroi,
  Phys.\ Rev.\ Lett.\  {\bf 84} (2000) 2322
  [hep-ph/9908309],
  Phys.\ Rev.\ D {\bf 61} (2000) 075005
  [hep-ph/9909334];
  J.~L.~Feng, K.~T.~Matchev and D.~Sanford,
  Phys.\ Rev.\ D {\bf 85} (2012) 075007
  [arXiv:1112.3021 [hep-ph]].

\bibitem{Abe:2007kf}
  H.~Abe, T.~Kobayashi and Y.~Omura,
  Phys.\ Rev.\ D {\bf 76} (2007) 015002
  [hep-ph/0703044];
  D.~Horton and G.~G.~Ross,
  Nucl.\ Phys.\ B {\bf 830} (2010) 221
  [arXiv:0908.0857 [hep-ph]];
  F.~Br\"ummer and W.~Buchm\"uller,
  JHEP {\bf 1107} (2011) 010
  [arXiv:1105.0802 [hep-ph]], JHEP {\bf 1205} (2012) 006
  [arXiv:1201.4338 [hep-ph]];
  J.~E.~Younkin and S.~P.~Martin,
  Phys.\ Rev.\ D {\bf 85} (2012) 055028
  [arXiv:1201.2989 [hep-ph]];
  T.~T.~Yanagida and N.~Yokozaki,
  Phys.\ Lett.\ B {\bf 722} (2013) 355
  [arXiv:1301.1137 [hep-ph]];
  F.~Br\"ummer, M.~Ibe and T.~T.~Yanagida,
  Phys.\  Lett.\ B {\bf 726} (2013) 364
  [arXiv:1303.1622 [hep-ph]].

\bibitem{Ellwanger:2009dp}
  U.~Ellwanger, C.~Hugonie and A.~M.~Teixeira,
  Phys.\ Rept.\  {\bf 496} (2010) 1
  [arXiv:0910.1785 [hep-ph]].

\bibitem{Batra:2003nj}
  P.~Batra, A.~Delgado, D.~E.~Kaplan and T.~M.~P.~Tait,
  JHEP {\bf 0402} (2004) 043
  [hep-ph/0309149].

\bibitem{Craig:2013cxa}
  N.~Craig,
  arXiv:1309.0528 [hep-ph].

\bibitem{Martin:1997ns}
  S.~P.~Martin,
  In *Kane, G.L. (ed.): Perspectives on supersymmetry II* 1-153
  [hep-ph/9709356].

\bibitem{Allanach:2001kg}
  B.~C.~Allanach,
  Comput.\ Phys.\ Commun.\  {\bf 143} (2002) 305
  [hep-ph/0104145].

\bibitem{Djouadi:2002ze}
  A.~Djouadi, J.~-L.~Kneur and G.~Moultaka,
  Comput.\ Phys.\ Commun.\  {\bf 176} (2007) 426
  [hep-ph/0211331].

\bibitem{Heinemeyer:1998yj}
  S.~Heinemeyer, W.~Hollik and G.~Weiglein,
  Comput.\ Phys.\ Commun.\  {\bf 124} (2000) 76
  [hep-ph/9812320],
  Eur.\ Phys.\ J.\ C {\bf 9} (1999) 343
  [hep-ph/9812472];
  G.~Degrassi, S.~Heinemeyer, W.~Hollik, P.~Slavich and G.~Weiglein,
  Eur.\ Phys.\ J.\ C {\bf 28} (2003) 133
  [hep-ph/0212020];
  M.~Frank, T.~Hahn, S.~Heinemeyer, W.~Hollik, H.~Rzehak and G.~Weiglein,
  JHEP {\bf 0702} (2007) 047
  [hep-ph/0611326].

\bibitem{Feng:2013tvd}
  J.~L.~Feng, P.~Kant, S.~Profumo and D.~Sanford,
  Phys.\ Rev.\ Lett.\  {\bf 111} (2013) 131802
  [arXiv:1306.2318 [hep-ph]].

\bibitem{Kant:2010tf}
  P.~Kant, R.~V.~Harlander, L.~Mihaila and M.~Steinhauser,
  JHEP {\bf 1008} (2010) 104
  [arXiv:1005.5709 [hep-ph]].


\bibitem{Pierce:1996zz}
  D.~M.~Pierce, J.~A.~Bagger, K.~T.~Matchev and R.~-j.~Zhang,
  Nucl.\ Phys.\ B {\bf 491} (1997) 3
  [hep-ph/9606211].

\bibitem{Csaki:2008sr}
  C.~Cs\'aki, A.~Falkowski, Y.~Nomura and T.~Volansky,
  Phys.\ Rev.\ Lett.\  {\bf 102} (2009) 111801
  [arXiv:0809.4492 [hep-ph]];
  A.~De Simone, R.~Franceschini, G.~F.~Giudice, D.~Pappadopulo and R.~Rattazzi,
  JHEP {\bf 1105} (2011) 112
  [arXiv:1103.6033 [hep-ph]].

\bibitem{Buchmuller:2007qf}
  W.~Buchm\"uller, C.~L\"udeling and J.~Schmidt,
  JHEP {\bf 0709} (2007) 113
  [arXiv:0707.1651 [hep-ph]].

\bibitem{Buchmuller:2005jr}
  W.~Buchm\"uller, K.~Hamaguchi, O.~Lebedev and M.~Ratz,
  Phys.\ Rev.\ Lett.\  {\bf 96} (2006) 121602
  [hep-ph/0511035].

\bibitem{Lebedev:2006kn}
  O.~Lebedev, H.~P.~Nilles, S.~Raby, S.~Ramos-Sanchez, M.~Ratz, P.~K.~S.~Vaudrevange and A.~Wingerter,
  Phys.\ Lett.\ B {\bf 645} (2007) 88
  [hep-th/0611095].

\bibitem{Hosteins:2009xk}
  P.~Hosteins, R.~Kappl, M.~Ratz and K.~Schmidt-Hoberg,
  JHEP {\bf 0907} (2009) 029
  [arXiv:0905.3323 [hep-ph]].

\bibitem{Lee:2003mc}
  H.~M.~Lee, H.~P.~Nilles and M.~Zucker,
  Nucl.\ Phys.\ B {\bf 680} (2004) 177
  [hep-th/0309195].

\bibitem{Buchmuller:2008uq}
  W.~Buchm\"uller and J.~Schmidt,
  Nucl.\ Phys.\ B {\bf 807} (2009) 265
  [arXiv:0807.1046 [hep-th]].

\bibitem{Buchmuller:2005rt}
  W.~Buchm\"uller, K.~Hamaguchi and J.~Kersten,
  Phys.\ Lett.\ B {\bf 632} (2006) 366
  [hep-ph/0506105].

\bibitem{ArkaniHamed:1998kj}
  N.~Arkani-Hamed, G.~F.~Giudice, M.~A.~Luty and R.~Rattazzi,
  Phys.\ Rev.\ D {\bf 58} (1998) 115005
  [hep-ph/9803290].

\bibitem{Chacko:1999mi}
  Z.~Chacko, M.~A.~Luty, A.~E.~Nelson and E.~Ponton,
  JHEP {\bf 0001} (2000) 003
  [hep-ph/9911323].

\bibitem{Krippendorf:2012ir}
  S.~Krippendorf, H.~P.~Nilles, M.~Ratz and M.~W.~Winkler,
  Phys.\ Lett.\ B {\bf 712} (2012) 87
  [arXiv:1201.4857 [hep-ph]];
  M.~Badziak, S.~Krippendorf, H.~P.~Nilles and M.~W.~Winkler,
  JHEP {\bf 1303} (2013) 094
  [arXiv:1212.0854 [hep-ph]].


\bibitem{Baer:2012uy}
  H.~Baer, V.~Barger, P.~Huang and X.~Tata,
  JHEP {\bf 1205} (2012) 109
  [arXiv:1203.5539 [hep-ph]];

\bibitem{Baer:2011ec}
  H.~Baer, V.~Barger and P.~Huang,
  JHEP {\bf 1111} (2011) 031
  [arXiv:1107.5581 [hep-ph]];
  M.~Berggren, F.~Br\"ummer, J.~List, G.~Moortgat-Pick, T.~Robens, K.~Rolbiecki and H.~Sert,
  arXiv:1307.3566 [hep-ph].

\bibitem{Atlas-sleptons} ATLAS collaboration, ATLAS-CONF-2013-049
\bibitem{CMS-sleptons} CMS collaboration, CMS-PAS-SUS-13-006

\bibitem{ArkaniHamed:1997ab}
  N.~Arkani-Hamed and H.~Murayama,
  Phys.\ Rev.\ D {\bf 56} (1997) 6733
  [hep-ph/9703259];
  F.~Br\"ummer, S.~Kraml and S.~Kulkarni,
  JHEP {\bf 1208} (2012) 089
  [arXiv:1204.5977 [hep-ph]];
  M.~Badziak, E.~Dudas, M.~Olechowski and S.~Pokorski,
  JHEP {\bf 1207} (2012) 155
  [arXiv:1205.1675 [hep-ph]].





\end{thebibliography}
\end{document}